\documentclass[conference,10pt]{IEEEtran}
\usepackage{url}
\usepackage[utf8]{inputenc}
\usepackage{xcolor}
\usepackage{amsmath}
\usepackage{amssymb}

\usepackage[acronyms,nonumberlist,nopostdot,nomain,nogroupskip]{glossaries}
\usepackage{tablefootnote}
\usepackage{booktabs}

\usepackage{tabularx}

\usepackage{tikz}
\usepackage{pgfplots}
\pgfplotsset{compat=newest} 
\pgfplotsset{plot coordinates/math parser=false} 
\newlength\fheight
\newlength\fwidth
\usetikzlibrary{plotmarks,patterns,decorations.pathreplacing,backgrounds,calc,arrows,arrows.meta,spy,matrix}
\usepgfplotslibrary{patchplots,groupplots}
\usepackage{tikzscale}
\usepackage{hyperref}

\newif\ifexttikz
\exttikzfalse

\ifexttikz
	\usetikzlibrary{external}
\fi

\usepackage{multirow}
\usepackage{tkz-kiviat}

\usepackage[font=scriptsize]{subcaption}
\usepackage[font=footnotesize]{caption}

\usepackage{mathtools}

\newacronym{3gpp}{3GPP}{3rd Generation Partnership Project}
\newacronym{adc}{ADC}{Analog to Digital Converter}
\newacronym{5g}{5G}{5th generation}
\newacronym{aimd}{AIMD}{Additive Increase Multiplicative Decrease}
\newacronym{am}{AM}{Acknowledged Mode}
\newacronym{amc}{AMC}{Adaptive Modulation and Coding}
\newacronym{aqm}{AQM}{Active Queue Management}
\newacronym{awgn}{AGWN}{Additive White Gaussian Noise}
\newacronym{balia}{BALIA}{Balanced Link Adaptation}
\newacronym{bdp}{BDP}{Bandwidth-Delay Product}
\newacronym{bf}{BF}{Beamforming}
\newacronym{cc}{CC}{Congestion Control}
\newacronym{cdf}{CDF}{Cumulative Distribution Function}
\newacronym{cn}{CN}{Core Network}
\newacronym{cqi}{CQI}{Channel Quality Information}
\newacronym{cp}{CP}{Control Plane}
\newacronym{csirs}{CSI-RS}{Channel State Information - Reference Signal}
\newacronym{dc}{DC}{Dual Connectivity}
\newacronym{dce}{DCE}{Direct Code Execution}
\newacronym{dci}{DCI}{Downlink Control Information}
\newacronym{dl}{DL}{Downlink}
\newacronym{dmr}{DMR}{Deadline Miss Ratio}
\newacronym{dmrs}{DMRS}{DeModulation Reference Signal}
\newacronym{e2e}{E2E}{End-to-End}
\newacronym{ecn}{ECN}{Explicit Congestion Notification}
\newacronym{edf}{EDF}{Earliest Deadline First}
\newacronym{enb}{eNB}{evolved Node Base}
\newacronym{epc}{EPC}{Evolved Packet Core}
\newacronym{es}{ES}{Edge Server}
\newacronym{fdma}{FDMA}{Frequency Division Multiple Access}
\newacronym{fdd}{FDD}{Frequency Division Duplexing}
\newacronym[firstplural=Radio Access Technologies (RATs)]{rat}{RAT}{Radio Access Technology}
\newacronym{fs}{FS}{Fast Switching}
\newacronym{ftp}{FTP}{File Transfer Protocol}
\newacronym{gnb}{gNB}{Next Generation Node Base}
\newacronym{harq}{HARQ}{Hybrid Automatic Repeat reQuest}
\newacronym{hetnet}{HetNet}{Heterogeneous Network}
\newacronym{hh}{HH}{Hard Handover}
\newacronym{hol}{HOL}{Head-of-Line}
\newacronym{ia}{IA}{Initial Access}
\newacronym{imt}{IMT}{International Mobile Telecommunication}
\newacronym{iot}{IoT}{Internet of Things}
\newacronym{los}{LOS}{Line-of-Sight}
\newacronym{lte}{LTE}{Long Term Evolution}
\newacronym{m2m}{M2M}{Machine to Machine}
\newacronym{mac}{MAC}{Medium Access Control}
\newacronym{mc}{MC}{Multi-Connectivity}
\newacronym{mcs}{MCS}{Modulation and Coding Scheme}
\newacronym{mec}{MEC}{Mobile Edge Cloud}
\newacronym{mi}{MI}{Mutual Information}
\newacronym{mimo}{MIMO}{Multiple Input, Multiple Output}
\newacronym{mmwave}{mmWave}{millimeter wave}
\newacronym{mptcp}{MPTCP}{Multipath TCP}
\newacronym{mr}{MR}{Maximum Rate}
\newacronym{mss}{MSS}{Maximum Segment Size}
\newacronym{mtd}{MTD}{Machine-Type Device}
\newacronym{mtu}{MTU}{Maximum Transmission Unit}
\newacronym{nfv}{NFV}{Network Function Virtualization}
\newacronym{nlos}{NLOS}{Non-Line-of-Sight}
\newacronym{nr}{NR}{New Radio}
\newacronym{ofdm}{OFDM}{Orthogonal Frequency Division Multiplexing}
\newacronym{pdcch}{PDCCH}{Physical Downlonk Control Channel}
\newacronym{pdcp}{PDCP}{Packet Data Convergence Protocol}
\newacronym{pdsch}{PDSCH}{Physical Downlink Shared Channel}
\newacronym{pdu}{PDU}{Packet Data Unit}
\newacronym{pf}{PF}{Proportional Fair}
\newacronym{pgw}{PGW}{Packet Gateway}
\newacronym{phy}{PHY}{Physical}
\newacronym{pbch}{PBCH}{Physical Broadcast Channel}
\newacronym[plural=\gls{mme}s,firstplural=Mobility Management Entities (MMEs)]{mme}{MME}{Mobility Management Entity}
\newacronym{prb}{PRB}{Physical Resource Block}
\newacronym{pss}{PSS}{Primary Synchronization Signal}
\newacronym{pucch}{PUCCH}{Physical Uplink Control Channel}
\newacronym{pusch}{PUSCH}{Physical Uplink Shared Channel}
\newacronym{rach}{RACH}{Random Access Channel}
\newacronym{ran}{RAN}{Radio Access Network}
\newacronym{red}{RED}{Random Early Detection}
\newacronym{rf}{RF}{Radio Frequency}
\newacronym{rlc}{RLC}{Radio Link Control}
\newacronym{rlf}{RLF}{Radio Link Failure}
\newacronym{rrc}{RRC}{Radio Resource Control}
\newacronym{rrm}{RRM}{Radio Resource Management}
\newacronym{rr}{RR}{Round Robin}
\newacronym{rs}{RS}{Remote Server}
\newacronym{rsrp}{RSRP}{Reference Signal Received Power}
\newacronym{rss}{RSS}{Received Signal Strength}
\newacronym{rtt}{RTT}{Round Trip Time}
\newacronym{rw}{RW}{Receive Window}
\newacronym{rx}{RX}{Receiver}
\newacronym{sa}{SA}{standalone}
\newacronym{sack}{SACK}{Selective Acknowledgment}
\newacronym{sap}{SAP}{Service Access Point}
\newacronym{sch}{SCH}{Secondary Cell Handover}
\newacronym{scoot}{SCOOT}{Split Cycle Offset Optimization Technique}
\newacronym{sdma}{SDMA}{Spatial Division Multiple Access}
\newacronym{sinr}{SINR}{Signal to Interference plus Noise Ratio}
\newacronym{sm}{SM}{Saturation Mode}
\newacronym{snr}{SNR}{Signal-to-Noise-Ratio}
\newacronym{son}{SON}{Self-Organizing Network}
\newacronym{ss}{SS}{Synchronization Signal}
\newacronym{srs}{SRS}{Sounding Reference Signal}
\newacronym{sss}{SSS}{Secondary Synchronization Signal}
\newacronym{tb}{TB}{Transport Block}
\newacronym{tcp}{TCP}{Transmission Control Protocol}
\newacronym{tdd}{TDD}{Time Division Duplexing}
\newacronym{tdma}{TDMA}{Time Division Multiple Access}
\newacronym{tfl}{TfL}{Transport for London}
\newacronym{tm}{TM}{Transparent Mode}
\newacronym{trp}{TRP}{Transmitter Receiver Pair}
\newacronym{tti}{TTI}{Transmission Time Interval}
\newacronym{ttt}{TTT}{Time-to-Trigger}
\newacronym{tx}{TX}{Transmitter}
\newacronym{ue}{UE}{User Equipment}
\newacronym{ul}{UL}{Uplink}
\newacronym{uml}{UML}{Unified Modeling Language}
\newacronym{um}{UM}{Unacknowledged Mode}
\newacronym{utc}{UTC}{Urban Traffic Control}
\newacronym{vm}{VM}{Virtual Machine}
\newacronym{rsrq}{RSRQ}{Reference Signal Received Quality}
\newacronym{rssi}{RSSI}{Received Signal Strength Indicator}
\newacronym{crs}{CRS}{Cell Reference Signal}
\newacronym{nsa}{NSA}{Non Stand Alone}
\newacronym{mrdc}{MR-DC}{Multi \gls{rat} \gls{dc}}
\newacronym{endc}{EN-DC}{E-UTRAN-\gls{nr} \gls{dc}}
\newacronym{5gc}{5GC}{5G Core}
\newacronym{si}{SI}{Study Item}
\newacronym{iab}{IAB}{Integrated Access and Backhaul}
\newacronym{wf}{WF}{Wired-first}
\newacronym{hqf}{HQF}{Highest-quality-first}
\newacronym{pa}{PA}{Position-aware}
\newacronym{mlr}{MLR}{Maximum-local-rate}
\newacronym{wbf}{WBF}{Wired Bias Function}
\newacronym{mib}{MIB}{Master Information Block}
\newacronym{sib}{SIB}{Secondary Information Block}
\newacronym{kpi}{KPI}{Key Performance Indicator}
\newacronym{ppp}{PPP}{Poisson Point Process}
\tikzstyle{startstop} = [rectangle, rounded corners, minimum width=2cm, minimum height=0.5cm,text centered, draw=black]
\tikzstyle{io} = [trapezium, trapezium left angle=70, trapezium right angle=110, minimum width=3cm, minimum height=1cm, text centered, draw=black]
\tikzstyle{process} = [rectangle, minimum width=2cm, minimum height=0.5cm, text centered, draw=black, alignb=center]
\tikzstyle{decision} = [ellipse, minimum width=2cm, minimum height=1cm, text centered, draw=black]
\tikzstyle{arrow} = [thick,<->,>=stealth]
\tikzstyle{line} = [thick,>=stealth]
\tikzstyle{darrow} = [thick,<->,>=stealth,dashed]
\tikzstyle{sarrow} = [thick,->,>=stealth]
\tikzstyle{larrow} = [line width=0.1mm,dashdotted,->,>=stealth]

\makeatletter
\def\grd@save@target#1{%
  \def\grd@target{#1}}
\def\grd@save@start#1{%
  \def\grd@start{#1}}
\tikzset{
  grid with coordinates/.style={
    to path={%
      \pgfextra{%
        \edef\grd@@target{(\tikztotarget)}%
        \tikz@scan@one@point\grd@save@target\grd@@target\relax
        \edef\grd@@start{(\tikztostart)}%
        \tikz@scan@one@point\grd@save@start\grd@@start\relax
        \draw[minor help lines] (\tikztostart) grid (\tikztotarget);
        \draw[major help lines] (\tikztostart) grid (\tikztotarget);
        \grd@start
        \pgfmathsetmacro{\grd@xa}{\the\pgf@x/1cm}
        \pgfmathsetmacro{\grd@ya}{\the\pgf@y/1cm}
        \grd@target
        \pgfmathsetmacro{\grd@xb}{\the\pgf@x/1cm}
        \pgfmathsetmacro{\grd@yb}{\the\pgf@y/1cm}
        \pgfmathsetmacro{\grd@xc}{\grd@xa + \pgfkeysvalueof{/tikz/grid with coordinates/major step x}}
        \pgfmathsetmacro{\grd@yc}{\grd@ya + \pgfkeysvalueof{/tikz/grid with coordinates/major step y}}
        \foreach \x in {\grd@xa,\grd@xc,...,\grd@xb}
        \node[anchor=north] at (\x,\grd@ya) {\pgfmathprintnumber{\x}};
        \foreach \y in {\grd@ya,\grd@yc,...,\grd@yb}
        \node[anchor=east] at (\grd@xa,\y) {\pgfmathprintnumber{\y}};
      }
    }
  },
  minor help lines/.style={
    help lines,
    gray,
    line cap =round,
    xstep=\pgfkeysvalueof{/tikz/grid with coordinates/minor step x},
    ystep=\pgfkeysvalueof{/tikz/grid with coordinates/minor step y}
  },
  major help lines/.style={
    help lines,
    line cap =round,
    line width=\pgfkeysvalueof{/tikz/grid with coordinates/major line width},
    xstep=\pgfkeysvalueof{/tikz/grid with coordinates/major step x},
    ystep=\pgfkeysvalueof{/tikz/grid with coordinates/major step y}
  },
  grid with coordinates/.cd,
  minor step x/.initial=.5,
  minor step y/.initial=.2,
  major step x/.initial=1,
  major step y/.initial=1,
  major line width/.initial=1pt,
}
\makeatother

\definecolor{desireRed}{RGB}{230,57,60}%
\definecolor{darkPurple}{RGB}{59,31,43}%
\definecolor{springGreen}{RGB}{37,223,145}%
\definecolor{queenBlue}{RGB}{69,123,157}%
\definecolor{spaceCadet}{RGB}{29,53,87}%

\makeglossaries

\begin{document}
	
\title{Distributed Path Selection Strategies\\for Integrated Access and Backhaul at mmWaves\vspace{-.3cm}}

\author{\IEEEauthorblockN{Michele Polese$^{\circ }$, Marco Giordani$^{\circ }$, Arnab Roy$^{\dagger }$, Douglas Castor$^{\dagger }$, Michele Zorzi$^{\circ }$}
\IEEEauthorblockA{
\small email:\texttt{\{polesemi,giordani,zorzi\}@dei.unipd.it, }
\texttt{\small \{arnab.roy,douglas.castor\}@interdigital.com}\\
$^{\circ }$\small Consorzio Futuro in Ricerca (CFR) and  University of Padova, Italy\\
$^{\dagger }$\small InterDigital Communications, Inc., USA\\}}

\makeatletter
\patchcmd{\@maketitle}
  {\addvspace{0.5\baselineskip}\egroup}
  {\addvspace{-1\baselineskip}\egroup}
  {}
  {}
\makeatother

\flushbottom
\setlength{\parskip}{0ex plus0.1ex}

\maketitle

\glsunset{nr}

\begin{abstract}
The communication at mmWave frequencies is a promising enabler for ultra high data rates in the next generation of mobile cellular networks (5G). The harsh propagation environment at such high frequencies, however, demands a dense base station deployment, which may be infeasible because of the unavailability of fiber drops to provide wired backhauling. To address this issue, 3GPP has recently proposed a Study Item on \gls{iab}, i.e., on the possibility of providing the wireless backhaul together with the radio access to the mobile terminals. 
The design of \gls{iab} base stations and networks introduces new research challenges, especially when considering the demanding conditions at mmWave frequencies. In this paper we study different path selection techniques, using a distributed approach, and investigate their performance in terms of hop count and bottleneck \gls{snr} using a channel model based on real measurements. We show that there exist solutions that decrease the number of hops without affecting the bottleneck \gls{snr} and provide guidelines on the design of \gls{iab} path selection policies.
\end{abstract}


\begin{picture}(0,0)(0,-360)
\put(0,0){
\put(0,0){\footnotesize This paper has been submitted to} 
\put(0,-10){\footnotesize IEEE Globecom 2018, Abu Dhabi, UAE}}
\end{picture}

\section{Introduction}

The next frontier of wireless communications is at mmWave frequencies~\cite{pi2011introduction}: the availability of large chunks of untapped spectrum makes it possible to exploit a large bandwidth and to satisfy the demand for multi-gigabit-per-second data rates in fifth generation (5G) cellular networks~\cite{ngmn5g}. The first field trials have confirmed the high potential of this technology~\cite{shafi2017deployment}, and for the first time 3GPP is standardizing mobile networks at mmWave frequencies. The new 5G \gls{ran} specifications (i.e., 3GPP \gls{nr}) will indeed support carrier frequencies up to 52.6 GHz, with a bandwidth per single carrier up to 400 MHz~\cite{38300}. 

The communication at such high frequencies, however, introduces new challenges and issues that must be taken into account when designing and deploying mmWave cellular networks~\cite{rangan2017potentials}. The first is related to the high propagation loss, which is proportional to the square of the carrier frequency for a given effective antenna area, and limits the communication range of mmWave base stations. Nonetheless, the short mmWave wavelength makes it possible to pack a large number of antenna elements in a limited area, thus enabling directional transmissions with a high beamforming gain that makes up for the high propagation loss~\cite{sun2014MIMO}. Therefore, 3GPP \gls{nr} supports directional transmissions in its \gls{phy} and \gls{mac} layer specifications, with procedures for a directional initial access and tracking of the best beam pair between users and base stations~\cite{giordani2018initial,38211}. The second issue is related to the blockage by common materials, such as brick, mortar and the human body~\cite{lu2012modeling}. While \gls{nlos} communications are possible also at mmWave frequencies~\cite{rappaport2013millimeter}, the presence of obstacles reduces the received signal power by up to three orders of magnitude, thus greatly increasing the probability of being in outage.

The combination of the high propagation loss and the blockage phenomenon advocates for a high-density deployment. With such network architecture, mmWave base stations would be deployed as small cells. Despite the limited coverage range, each single mobile terminal would be in \gls{los} with respect to multiple links, at any given time, and therefore its outage probability would decrease~\cite{rangan2017potentials}. Such ultra-dense deployment, however, can be costly for network operators~\cite{lopezperez2015towards}, because of the capital and operational expenditures and the need for a reliable and high capacity backhaul between the base stations and the operator core network. This issue motivated the approval of a new \gls{si} for 3GPP Release 15~\cite{iabsi2017}, which will analyze the feasibility of an \gls{iab} deployment for \gls{nr}, in which the backhaul links are wireless and operated in the same context of the \gls{ran} access. 

In this paper we present novel results related to the choice of the backhaul path in an \gls{iab} setup, using a mmWave channel based on real measurements, with a realistic beamforming model and a sectorized deployment. We compare how different greedy policies perform with respect to the number of hops and the bottleneck \gls{snr}, i.e., the \gls{snr} of the weakest wireless backhaul link, relying only on local information, without the need for a centralized coordinator. Moreover, we discuss the usage of a function that biases the link selection towards base stations with a wired backhaul to the core network, and show that, for a certain set of parameters for this bias, it is possible to decrease the number of hops without affecting the average bottleneck \gls{snr}. This study can be used as a guideline for the choice and the design of backhaul path selection policies in \gls{iab} mmWave networks.  

The remainder of the paper is organized as follows. Sec.~\ref{sec:iab} describes the characteristics of the 3GPP \gls{si} on \gls{iab} and the potential of this solution for \gls{nr} deployments. Sec.~\ref{sec:policies} presents the link selection policies analyzed in this paper, while Sec.~\ref{sec:perf_eval} introduces the system model and the results of the performance evaluation. Finally, Sec.~\ref{sec:concl} concludes the paper and discusses possible extensions of this work.

\section{Integrated Access and Backhaul in 3GPP NR}
\label{sec:iab}

The research on wireless backhaul solutions has spanned the last two decades, with the goal of replacing costly fixed links with more flexible wireless connections. For example, mesh and multihop wireless backhaul architectures have been extensively studied for IEEE 802.11 networks~\cite{gambiroza2004multihop,alicherry2005joint}. However, in the cellular domain, integrated solutions that provide both access and backhaul functionalities have not been widely adopted yet. 
There exists a relay functionality integrated in the \gls{lte} specifications, which however has not been extensively deployed due to its limited flexibility~\cite{36300}: the resource configuration is fixed, it supports only single-hop relaying, and there is a fixed association between the relay and the parent base station that connects it to the wired core network.
On the other hand, the wireless backhaul links that are actually used to complement fiber optic cables for backhauling traffic in sub-6 GHz cellular networks are usually custom point-to-point solutions, not integrated with the \gls{ran}.

Nonetheless, the integration of the wireless backhaul with the radio access is being considered as a promising solution for 5G cellular networks. Papers~\cite{ge2014wireless,dehos2014millimeter} provided preliminary results on wireless backhaul for 5G, using also mmWave links, and showed that such solutions can meet the expected increase in mobile traffic demands. However, they did not consider a tight integration between the access and the backhaul, which is instead the focus of the more recent 3GPP \gls{si} on \gls{iab} for 3GPP \gls{nr}~\cite{iabsi2017}. The goal of this study item is to design an advanced wireless relay, which overcomes the limitations of the \gls{lte} relay, and makes it possible to deploy self-backhauled \gls{nr} base stations in a \textit{plug-and-play} manner.

According to~\cite{iabsi2017}, \gls{nr} cellular networks with \gls{iab} relays will be characterized by (i) the integration of the access and backhaul technologies; (ii) a higher flexibility in terms of network deployment and configuration with respect to \gls{lte}; and (iii) the possibility of using the mmWave spectrum. In particular, the design goal for \gls{iab} is to simplify both the installation and the management of dense \gls{nr} networks, exploiting the self-backhauling functionality integrated with the access, for example with plug-and-play \gls{iab} nodes capable of self-configuring and self-optimizing themselves~\cite{22261}. As stated in~\cite{iabsi2017,22261}, 5G \gls{iab} relays will be used in both outdoor and indoor scenarios, also with multiple wireless hops, in order to extend the coverage, and should be able to reconfigure the topology autonomously in order to avoid service unavailability. Moreover, a flexible split between the access and the backhaul resources is envisioned, in order to increase the efficiency of the resource allocation. Both in-band and out-of-band backhaul will be considered, with the first being a natural candidate for a tighter integration between access and backhaul. 

Furthermore, the usage of mmWave frequencies for \gls{iab} nodes introduces new opportunities and challenges. In particular, the directionality of mmWave links implies a higher spatial reuse, possibly enabling a spatial division multiple access scheme and a higher throughput, as discussed in~\cite{islam2017integrated}. On the other hand, the harsh propagation environment in the mmWave band requires a prompt adaptation of the topology and a fast link selection in case of outage, together with a dynamic scheduling process that adjusts the resource partition between access and backhaul according to the respective load. Therefore, mmWave \gls{iab} nodes can fully benefit from the flexibility and the self-organizing properties envisioned in the 3GPP \gls{si} for \gls{iab}.

In this regard, some papers recently analyzed the performance of \gls{iab} deployments at mmWaves, focusing however primarily on scheduling. In~\cite{yuan2018optimal}, the authors consider a centralized scheduling and routing problem, and show its performance in terms of throughput and computational run time required to find the optimal solution. Similarly, paper~\cite{niu2017energy} considers a joint optimization of the scheduling and the power, with the energy efficiency of the system as a target. In~\cite{saha2018integrated}, the authors focus on the resource split between access and backhaul, without considering link selection for \gls{iab} nodes. None of these works, however, considers a channel characterized by the full channel matrix, with large and small scale fading phenomena, nor realistic beamforming, in the performance evaluation.

\section{Path Selection Policies for \gls{iab} at mmWaves}
\label{sec:policies}
In this paper, we use the NYU channel model for mmWave frequencies described in~\cite{akdeniz2014millimeter} to analyze the performance of different path selection policies for the backhaul. In the following paragraphs, we will use the term (i) \textit{wired} \gls{gnb} or \textit{donor} to identify \glspl{gnb} which are connected to the core network with a wired backhaul; (ii) \gls{iab} node or relay to label \glspl{gnb} which do not have a wired backhaul link; and (iii) parent \gls{gnb} to name a \gls{gnb} which provides a wireless backhaul link to an \gls{iab} node. The parent can be itself a wireless \gls{iab} node, or a wired \gls{gnb}.

	\begin{table*}[t]
	\footnotesize
	\centering
	\def\tabularxcolumn#1{m{#1}}

	\begin{center}
	\renewcommand{\arraystretch}{1.3}

	\begin{tabularx}{0.98\textwidth}{>{\hsize=0.2\textwidth}l>{\hsize=0.1\textwidth}X>{\hsize=0.245\textwidth}X>{\hsize=0.245\textwidth}X>{\hsize=0.245\textwidth}X}

		\toprule
			\textbf{Policy}		&	Metric 		&	Selection rule 										& 	Pros 	& 	Cons \\ \midrule
		\textbf{HQF}  	&	\gls{snr}	&	Select the link with the highest \gls{snr} 	&	High bottleneck \gls{snr}	&	High probability of not reaching a wired \gls{gnb} \\

		\textbf{WF}  	&	\gls{snr}	&	Select the wired \gls{gnb}, if available, otherwise apply HQF &	Low number of hops	&	Low bottleneck \gls{snr} \\

		\textbf{PA}  	&	\gls{snr}	&	Select the link with the highest \gls{snr} among those with parents which are closer to a wired \gls{gnb} & Low number of hops	&	Possible ping-pong effects \\

		\textbf{MLR}  	&	Load and Shannon rate	&	Select the link with the highest achievable rate &	High bottleneck rate, traffic balancing	&	High probability of not reaching a wired \gls{gnb}\\

		\bottomrule
	\end{tabularx}
	\end{center}
	\caption{Comparison between the different link selection policies studied in this paper.}
	\label{table:policies}
\end{table*}

For all of the policies, the \gls{iab} node that has to find the path towards the core network initiates the procedure by applying the selection policy on the first hop, and then the procedure continues iteratively at each hop until a suitable wired \gls{gnb} is reached. Therefore, the strategies we evaluate are greedy, i.e., consider local information\footnote{With the exception of information related to the position and the backhaul technology, which can however be shared in advance.} to perform the hop-by-hop link selection decisions, and do not need a centralized controller. These policies can be used to re-route backhaul traffic on the fly, in case of a link failure, and to connect (possibly via multiple hops) an \gls{iab} node which is joining the network for the first time to a suitable wired \gls{gnb} in an autonomous and non-coordinated fashion. 

In Sec.~\ref{sec:policy}, we will describe each single strategy, while in Sec.~\ref{sec:bias} we will introduce the bias functions that we designed in order to improve the forwarding performance in terms of hops.

\subsection{Path Selection Policies}
\label{sec:policy}
The considered policies differ from one another because of the metric used to measure the link quality (\gls{snr} or rate), and because of the ranking criterion of the different available links at each hop. 
For every policy, and at each hop, we consider an \gls{snr} threshold $\Gamma_{\rm th}$, i.e., for the link selection, we compare only backhaul connections with an \gls{snr} $\Gamma$ higher than or equal to this threshold. If $\Gamma_{\rm th}$ is small, then it is possible to select and compare a larger number of base stations as parent candidates, and possibly increase the probability of successfully reaching a wired \gls{gnb}, at the price of a lower data rate on the bottleneck link.
For the access network, $\Gamma_{\rm th}$ is usually set to $-5$ dB~\cite{giordani2018initial}, i.e., access links with an \gls{snr} smaller than $-5$ dB are usually considered in outage. However, this choice is not valid in a backhaul context, where the link is required to reliably forward high-data-rate traffic from the relay to its parent \gls{gnb}. Therefore, we  select a higher value for $\Gamma_{\rm th}$, i.e., 5~dB, which corresponds to a theoretically achievable Shannon rate of 830 Mbit/s, on a single carrier with a bandwidth $B = 400$~MHz~\cite{38211}. Moreover, we avoid loops, i.e., if an \gls{iab} node was used as a relay in a previous hop, it cannot be selected  again.

Table~\ref{table:policies} sums up the main properties of each policy, which are described in detail in the following paragraphs.

\paragraph*{\gls{hqf} policy} At each hop, the \gls{hqf} strategy compares the \gls{snr} $\Gamma$ of the available links towards each possible parent \glspl{gnb} (either wired or wireless), and selects that with the highest \gls{snr}, without considering any additional information. It is a very simple selection rule, which can be implemented only by measuring the link quality using synchronization signals. Moreover, by always selecting the best \gls{snr}, the bottleneck link, i.e., the link with the lowest \gls{snr} among the hops towards the wired \gls{gnb}, will have a high \gls{snr} when compared to other policies. On the other hand, given that this policy follows a greedy approach, it may happen that the parent \gls{gnb} with the best \gls{snr} leads further away from a wired \gls{gnb}, thus increasing the number of hops. Moreover, in some cases, the highest \gls{snr} leads to the choice of another relay \gls{gnb} which however is not within reach of any other possible wireless parent or wired donor, thus failing to connect to a wired \gls{gnb}. 

\paragraph*{\gls{wf} policy} The \gls{wf} policy is designed to reduce as much as possible the number of hops needed to reach a wired \gls{gnb}. Indeed, if at a given hop one of the available backhaul links is toward a wired \gls{gnb}, i.e., if a wired \gls{gnb} is reachable from the current \gls{iab} node with an \gls{snr} higher than the threshold $\Gamma_{\rm th}$, then the wired \gls{gnb} is selected even if it is not associated to the connection with the highest \gls{snr}. If instead no wired \gls{gnb} is available, then the \gls{hqf} policy is applied. The \gls{iab} node would need to know which candidate parents are wired or wireless, and this can be done by extending the information directionally broadcasted (using \gls{ss} blocks~\cite{38331}) by each \gls{gnb} in the \gls{mib} or \gls{sib}.
While this policy increases the probability of reaching a wired \gls{gnb}, even with a greedy approach, it may cause a degradation in the quality of the bottleneck link.

\paragraph*{\gls{pa} policy} This strategy uses additional context information related to the position of the \gls{iab} node that has to perform the link selection and the wired \gls{gnb} in the scenario. This information can be available in advance and pre-configured in the relays (especially if non-mobile relays are considered~\cite{iabsi2017}), or shared on directional broadcast messages. The goal is to avoid selecting a parent \gls{gnb} that is more distant from the closest wired \gls{gnb} than the current \gls{iab} node. Therefore, the \gls{iab} node divides the neighboring region into two half-planes, identified by the line which is perpendicular to the one that passes through the \gls{iab} and the wired \glspl{gnb} positions. Then, it considers for its selection only the candidate parents which are in the half-plane with also the wired \gls{gnb}, and selects that with the highest \gls{snr}. This policy should strike a balance between \gls{hqf} and \gls{wf}.

\paragraph*{\gls{mlr} policy} The \gls{mlr} policy does not consider the \gls{snr} as a metric, but at each hop selects the candidate parent with the highest achievable Shannon rate. Consider \gls{iab} node $i$, and the candidate parent $j$, with $N_j$ among users and \gls{iab} nodes attached. Then, given a bandwidth $B$ and the \gls{snr} $\Gamma_{i, j}$ between the \gls{iab} node and the candidate parent, the Shannon rate is computed as $R_j = B / N_j \log_2 (1 + \Gamma_{i,j})$. Finally, the \gls{iab} node selects the parent with the highest achievable rate $R$. Once again, we assume that the information on the load (in terms of number of users $N_j$) of candidate parent $j$ is known to the \gls{iab} node, for example through extension of the \gls{mib} or \gls{sib}, or with a passive estimation of the power ratio between the resources allocated to synchronization signals and data transmissions. This strategy is designed to take into account the load information in the decision, but has the same drawbacks of the \gls{hqf} policy, i.e., it may yield a high number of hops and/or  connection failures.

\subsection{Wired Bias Function}
\label{sec:bias}
For multi-hop scenarios, one of the \acrlongpl{kpi} that is considered in the 3GPP \gls{si} for \gls{iab} is the number of hops from a certain wireless \gls{iab} node to the first wired \gls{gnb} it can reach. However, as discussed in the previous section, some of the proposed policies may need a high number of hops, or even never reach the target wired \gls{gnb}. In order to solve this issue, it is possible to apply a \gls{wbf} to the \gls{snr} of the wired \glspl{gnb} during the evaluation of the metric for the link selection. Consequently, a wired \gls{gnb} may be chosen as parent even though it is not the candidate with the highest considered metric. 

The bias is not fixed, but is a function $W(N)$ of the number of hops $N$ traveled from the \gls{iab} node that is trying to connect to a wired \gls{gnb}. The idea is that as $N$ increases, it becomes more and more convenient to select as a parent a wired \gls{gnb} with respect to another wireless \gls{iab} node (that would otherwise add up to the number of hops) even though the wired \gls{gnb} is not the best according to the metric considered. 
The \gls{wf} policy is a particular case of a decision with bias, with $W(N)$ large enough so that the wired \gls{gnb} is always selected if above the $\Gamma_{\rm th}$ threshold.   

We compare two different \glspl{wbf}, which are respectively polynomial and exponential in the number of hops $N$. The first is defined as follows:
\begin{equation}
	W_{p}(N) = \left ( \frac{N}{N_{h,t}} \right )^k \Gamma_{gap} + \Gamma_H,
		\label{eq:Wp}
\end{equation}
where $k$ is the degree of the polynomial, $N_{h,t}$ is a threshold on the number of hops, $\Gamma_{gap}$ a tolerable \gls{snr} gap, and $\Gamma_H$ an \gls{snr} hysteresis. The idea is that, if $N$ is smaller than $N_{h,t}$, then the \gls{snr} gap parameter $\Gamma_{gap}$ is multiplied by a number smaller than 1, and the \gls{wbf} $W(N)$ does not impact too much the link choice. When the number of hops $N$ reaches the threshold $N_{h,t}$, then $W(N)$ assumes values which are greater than or equal to $\Gamma_{gap}$, increasing the weight of the bias in the link selection. The \gls{snr} hysteresis $\Gamma_H$ is set to 2~dB, and slightly offsets the choice towards a wired \gls{gnb} in case the best wireless relay candidate and the wired \gls{gnb} have a very similar \gls{snr}. Very conservative \gls{wbf} would use a large $N_{h,t}$, and a small $k$ and $\Gamma_{gap}$, and vice versa for an aggressive parameter tuning. 

Similarly, the exponential \gls{wbf} is  defined as
\begin{equation}
	W_{e}(N) = \gamma^{\left ( \frac{N}{N_{h,t}} \right )} \Gamma_{gap} + \Gamma_H,
	\label{eq:We}
\end{equation}
with $\gamma$ the basis of the exponential function. Notice that $\gamma$ must be greater than or equal to 1, otherwise $\gamma^{\left ( \frac{N}{N_{h,t}}\right )}$ would decrease with the number of hops. Moreover, for any $\gamma$, the exponential \gls{wbf} $W_{e}(N)$ is larger than the polynomial $W_{p}(N)$, for the same choice of the other parameters. For example, if $N_{h,t} = 6$ and $N=1$, with $\gamma = 1.5$ we have $\gamma^{\left ( \frac{N}{N_{h,t}}\right )} = 1.07$, while with $k=1$ we have $\left ( \frac{N}{N_{h,t}} \right )^k = 0.17$.

\section{Performance Evaluation}
\label{sec:perf_eval}

In this section, we first provide some details on the system model used for the performance evaluation and then discuss the simulation results and compare the different policies described in Sec.~\ref{sec:policies}.

\subsection{System Model}
\label{sec:system_model}

The performance evaluation for this paper is done with Monte Carlo simulations with 20000 independent repetitions for each single configuration. The main parameters for the simulations are reported in Table~\ref{tab:params}.

The \glspl{gnb} (both wired and wireless) are deployed with a \gls{ppp} with density $\lambda_g \in \{30, 60\}$ \gls{gnb}/km$^2$, and a fraction $p_w \in \{0.1, 0.3\}$ is configured with a wired backhaul link to the core network. Therefore, the density of the wired \glspl{gnb} is $\lambda_{w,g} = p_w\lambda_g$ \gls{gnb}/km$^2$, while the \gls{iab} nodes have a density $\lambda_{i,g} = (1 - p_{w}) \lambda_g$ \gls{gnb}/km$^2$. For the evaluation of the \gls{mlr} policy, we also deploy \glspl{ue} with a \gls{ppp} and a density of $\lambda_{\rm UE}$ \gls{ue}/km$^2$. They are associated to the \gls{gnb} with the smallest pathloss, in line with previous studies~\cite{saha2018integrated}.

We assume that the \gls{iab} are equipped with $S$ uniform planar antenna arrays, with the same number $M \in \{64, 256\}$ of isotropic antenna elements at both endpoints of the connection. Each antenna array covers a sector of $2\pi/S$ degrees. Moreover, node $i$ can monitor the link quality of the neighboring \gls{gnb} $j \in \mathcal{N}_i$, where $\mathcal{N}_i$ is the set of wired or wireless \glspl{gnb} whose reference signals can be received by node $i$. The \gls{iab} node can then select the best beam to communicate with $j$ using the standard beam management procedures of 3GPP \gls{nr}\footnote{One of the goals of the \gls{iab} \gls{si}, indeed, is to reuse the \gls{nr} specifications for the access links also for the backhaul. In any case, enhancements related to the backhaul functionality can be introduced, thanks to the more advanced capabilities of an \gls{iab} node with respect to a mobile \gls{ue}~\cite{iabsi2017}.}. 

Table~\ref{table:wbf} summarizes the main parameters used for the \gls{wbf}. In particular, we identify a conservative policy, with $N_{h,t} = 6$, $\Gamma_{gap} = 5$ dB and $k=1$ or $\gamma = 1.5$ for the polynomial and the exponential policies, respectively, and an aggressive one, with $N_{h,t} = 1$, $\Gamma_{gap} = 15$ dB and $k=3$ or $\gamma = 3$. 

\begin{table}[t]
\vspace{0.05in}
\renewcommand{\arraystretch}{1}
\footnotesize
\centering
\begin{tabular}{@{}lll@{}}
\toprule
Parameter & Value & Description \\ \midrule
$B$ & 400 MHz & Bandwidth of mmWave \glspl{gnb}\\
$f_{\rm c}$ & $28$ GHz & mmWave carrier frequency \\
$P_{\rm TX}$ & $30$ dBm & mmWave transmission power \\
NF  & $5$ dB & Noise figure \\
$M$ & $\{8 \times 8, 16 \times 16\}$  & \gls{gnb} UPA MIMO array size  \\
$S$ & 3 & Number of sectors for each \gls{gnb} \\
$\lambda_g$ & $\{30, 60\}$ gNB/km$^2$ & \gls{gnb} density \\
$p_w$ & $\{0.1, 0.3\}$ & Fraction of wired \gls{gnb} \\
\bottomrule
\end{tabular}
\caption{Simulation parameters.}
\label{tab:params}
\end{table}

\begin{table}
\renewcommand{\arraystretch}{1}
\footnotesize
\centering
\begin{tabular}{@{}ll@{}}
\toprule
Configuration & Parameters \\\midrule
Aggressive $W_{e}(N)$ & $N_{h,t} = 1$, $\gamma=3$, $\Gamma_{gap} = 15$ dB, $\Gamma_H=2$ dB \\
Conservative $W_{e}(N)$ & $N_{h,t} = 6$, $\gamma=1.5$, $\Gamma_{gap} = 5$ dB, $\Gamma_H=2$ dB \\
Aggressive $W_{p}(N)$ & $N_{h,t} = 1$, $k=3$, $\Gamma_{gap} = 15$ dB, $\Gamma_H=2$ dB \\
Conservative $W_{p}(N)$ & $N_{h,t} = 6$, $k=1$, $\Gamma_{gap} = 5$ dB, $\Gamma_H=2$ dB \\
\bottomrule
\end{tabular}
\caption{\acrlong{wbf} parameters.%
}
\label{table:wbf}
\end{table}

\subsection{Results and Discussion}
\label{sec:results}

The performance of the  \gls{iab} path selection schemes will be evaluated by comparing the \glspl{cdf} of (i) the number of hops required to forward the backhaul traffic from a wireless to a wired \gls{gnb}, and (ii) the bottleneck \gls{snr}, i.e., the \gls{snr} of the weakest link.

\ifexttikz
	\tikzsetnextfilename{cdfHopsWired30percentDiffLambdaAntenna}
\fi
\begin{figure}[t]
	\centering
	\setlength\fwidth{0.8\columnwidth}
	\setlength\fheight{0.5\columnwidth}
%
%
%
\begin{tikzpicture}
\pgfplotsset{every tick label/.append style={font=\scriptsize}}
\begin{axis}[%
width=\fwidth,
height=\fheight,
at={(0\fwidth,0\fheight)},
scale only axis,
unbounded coords=jump,
axis y line*=left,
xmin=0.9,
xmax=8,
xlabel style={font=\scriptsize\color{white!15!black}},
xlabel={Number of hops to reach the wired gNB},
ymin=0,
ymax=1.02,
ylabel style={font=\scriptsize\color{white!15!black}},
ylabel={CDF},
axis background/.style={fill=white},
xmajorgrids,
ymajorgrids,
legend style={font=\scriptsize,legend cell align=left, align=left, draw=white!15!black,at={(0.5, 1.03)},anchor=south},
legend columns=2,
xlabel shift=-2pt,
ylabel shift=-2pt,
xticklabel shift=-1pt,
yticklabel shift=-1pt,
]
\addplot [color=desireRed, line width=0.9pt, forget plot]
  table[row sep=crcr, x expr=\thisrow{X}+1]{%
X	Y\\
-inf	0\\
0	0\\
0	0.732058995252964\\
1	0.732058995252964\\
1	0.884408397431016\\
2	0.884408397431016\\
2	0.949508795978981\\
3	0.949508795978981\\
3	0.978511410656716\\
4	0.978511410656716\\
4	0.991305561901861\\
5	0.991305561901861\\
5	0.996484146929658\\
6	0.996484146929658\\
6	0.998946513339933\\
7	0.998946513339933\\
7	0.999619221689133\\
8	0.999619221689133\\
8	0.999923844337827\\
9	0.999923844337827\\
9	0.999961922168913\\
10	0.999961922168913\\
10	0.999987307389638\\
11	0.999987307389638\\
11	1\\
inf	1\\
};

\addplot [color=desireRed, line width=0.9pt, forget plot, mark=x, mark size=1.5pt, only marks]
  table[row sep=crcr, x expr=\thisrow{X}+1]{%
X	Y\\
0.5	0.732058995252964\\
1.5	0.884408397431016\\
2.5	0.949508795978981\\
3.5	0.978511410656716\\
4.5	0.991305561901861\\
5.5	0.996484146929658\\
6.5	0.998946513339933\\
7.5	0.999619221689133\\
8.5	0.999923844337827\\
9.5	0.999961922168913\\
10.5	0.999987307389638\\
};

\addplot [color=desireRed, line width=0.9pt, mark=x, mark size=1.5pt]
  table[row sep=crcr, x expr=\thisrow{X}+1]{%
X	Y\\
0 -1 \\
}; 
\addlegendentry{$M = 64, \lambda_g = 30$ gNB/km$^2$}

\addplot [color=desireRed, dashdotted, line width=0.9pt]
  table[row sep=crcr, x expr=\thisrow{X}+1]{%
X	Y\\
-inf	0\\
0	0\\
0	0.848829014697321\\
1	0.848829014697321\\
1	0.944121582285592\\
2	0.944121582285592\\
2	0.977540369118263\\
3	0.977540369118263\\
3	0.990511597503805\\
4	0.990511597503805\\
4	0.996118845049995\\
5	0.996118845049995\\
5	0.998273907403814\\
6	0.998273907403814\\
6	0.999397399626183\\
7	0.999397399626183\\
7	0.999826369383816\\
8	0.999826369383816\\
8	0.999928505040395\\
9	0.999928505040395\\
9	0.999979572868684\\
10	0.999979572868684\\
10	0.999989786434342\\
12	0.999989786434342\\
12	1\\
inf	1\\
};
\addlegendentry{$M = 64, \lambda_g = 60$ gNB/km$^2$}

\addplot [color=darkPurple, line width=0.9, forget plot]
  table[row sep=crcr, x expr=\thisrow{X}+1]{%
X	Y\\
-inf	0\\
0	0\\
0	0.756237513357803\\
1	0.756237513357803\\
1	0.889966547414394\\
2	0.889966547414394\\
2	0.949426195233007\\
3	0.949426195233007\\
3	0.978906286298379\\
4	0.978906286298379\\
4	0.991125772429494\\
5	0.991125772429494\\
5	0.996480509222692\\
6	0.996480509222692\\
6	0.998571295823073\\
7	0.998571295823073\\
7	0.999454072387678\\
8	0.999454072387678\\
8	0.999767690377735\\
9	0.999767690377735\\
9	0.999941922594434\\
10	0.999941922594434\\
10	0.999988384518887\\
13	0.999988384518887\\
13	1\\
inf	1\\
};

\addplot [color=darkPurple, line width=0.9, forget plot, mark=o, mark size=1pt, only marks]
  table[row sep=crcr, x expr=\thisrow{X}+1]{%
X	Y\\
-inf	0\\
0.5	0.756237513357803\\
1.5	0.889966547414394\\
2.5	0.949426195233007\\
3.5	0.978906286298379\\
4.5	0.991125772429494\\
5.5	0.996480509222692\\
6.5	0.998571295823073\\
7.5	0.999454072387678\\
8.5	0.999767690377735\\
9.5	0.999941922594434\\
10.5	0.999988384518887\\
};

\addplot [color=darkPurple, line width=0.9, mark=o, mark size=1pt]
  table[row sep=crcr, x expr=\thisrow{X}+1]{%
X	Y\\
0	-1\\
};
\addlegendentry{$M = 256, \lambda_g = 30$ gNB/km$^2$}

\addplot [color=darkPurple, dashdotted, line width=0.9pt]
  table[row sep=crcr, x expr=\thisrow{X}+1]{%
X	Y\\
-inf	0\\
0	0\\
0	0.894595112505169\\
1	0.894595112505169\\
1	0.961775473772327\\
2	0.961775473772327\\
2	0.985204385230608\\
3	0.985204385230608\\
3	0.994200764490525\\
4	0.994200764490525\\
4	0.997609706407399\\
5	0.997609706407399\\
5	0.999001522929673\\
6	0.999001522929673\\
6	0.999536061159242\\
7	0.999536061159242\\
7	0.999828544341459\\
8	0.999828544341459\\
8	0.999949571865135\\
9	0.999949571865135\\
9	0.999979828746054\\
10	0.999979828746054\\
10	0.999989914373027\\
15	0.999989914373027\\
15	1\\
inf	1\\
};
\addlegendentry{$M = 256, \lambda_g = 60$ gNB/km$^2$}

\end{axis}

\input{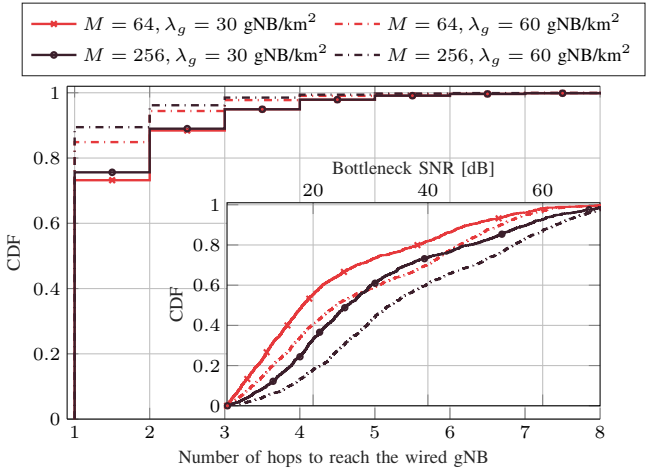}
\end{tikzpicture}%
	\caption{Performance of the \gls{wf} policy with different number of antennas $M$ and \gls{gnb} density $\lambda_g$.}
	\label{fig:wired_config}
\end{figure}

\textbf{Antenna and deployment configurations --}
In Fig.~\ref{fig:wired_config} we investigate how the relaying performance evolves as a function of different setup configurations, i.e., the number of antenna elements $M$ each \gls{gnb} is equipped with  and the gNB density $\lambda_g$.  The \gls{wf} strategy is considered.
As expected, increasing the MIMO array size has beneficial effects on both the number of hops and the bottleneck \gls{snr}.
In the first case,  the narrower beams that can be steered and the resulting  higher gains that are produced by beamforming enlarge the discoverable area of each \gls{gnb}, thereby increasing the probability of detecting a wired \gls{gnb} with sufficiently good signal quality and through a limited number of hops.
In the second case, sharper beams guarantee better signal quality and, consequently, stronger received power.

Similarly, enhanced backhauling performance is achieved by densifying the network since the \glspl{gnb} are gradually closer and thus establish more precise and, in general, more accurate communications.
Of course, if we persistently keep on increasing $\lambda_g$, the performance gain would progressively reduce  because of the increasingly higher impact of the interference from the surrounding base stations.

Finally, notice that the $M=64$, $\lambda_g=60 $ gNB/km$^2$ and the  $M=256$, $\lambda_g=30 $ gNB/km$^2$ configurations show, on average, comparable performance in terms of bottleneck \gls{snr}. However, for low \gls{snr} regimes, i.e., when considering farther nodes and more demanding signal propagation characteristics, densification is more effective than directionality.

\ifexttikz
	\tikzsetnextfilename{cdfHopsLambda30Antenna64DiffPolicies}
\fi
\begin{figure}[t]
	\centering
	\setlength\fwidth{0.8\columnwidth}
	\setlength\fheight{0.45\columnwidth}
%
%
%
\begin{tikzpicture}
\pgfplotsset{every tick label/.append style={font=\scriptsize}}
\begin{axis}[%
width=\fwidth,
height=\fheight,
at={(0\fwidth,0\fheight)},
scale only axis,
unbounded coords=jump,
axis y line*=left,
xmin=0.9,
xmax=11,
xlabel style={font=\scriptsize\color{white!15!black}},
xlabel={Number of hops to reach the wired gNB},
ymin=0,
ymax=1.02,
ylabel style={font=\scriptsize\color{white!15!black}},
ylabel={CDF},
axis background/.style={fill=white},
xmajorgrids,
ymajorgrids,
legend style={font=\scriptsize,legend cell align=left, align=left, draw=white!15!black,at={(0.5, 1.03)},anchor=south},
legend columns=3,
xlabel shift=-2pt,
ylabel shift=-2pt,
xticklabel shift=-1pt,
yticklabel shift=-1pt,
]
\addplot [color=spaceCadet, line width=0.9pt, forget plot]
  table[row sep=crcr, x expr=\thisrow{X}+1]{%
X	Y\\
-inf	0\\
0	0\\
0	0.512648184448801\\
1	0.512648184448801\\
1	0.690282833940073\\
2	0.690282833940073\\
2	0.797349015215159\\
3	0.797349015215159\\
3	0.87407822272006\\
4	0.87407822272006\\
4	0.924670960515262\\
5	0.924670960515262\\
5	0.95566134602819\\
6	0.95566134602819\\
6	0.974890320171754\\
7	0.974890320171754\\
7	0.986465042471763\\
8	0.986465042471763\\
8	0.99290581536451\\
9	0.99290581536451\\
9	0.996452907682255\\
10	0.996452907682255\\
10	0.997853075702418\\
11	0.997853075702418\\
11	0.999066554653225\\
12	0.999066554653225\\
12	0.99925324372258\\
13	0.99925324372258\\
13	0.999813310930645\\
14	0.999813310930645\\
14	0.999906655465323\\
15	0.999906655465323\\
15	1\\
inf	1\\
};

\addplot [color=spaceCadet, line width=0.9pt, mark=x, mark size=1.5pt, only marks, forget plot]
  table[row sep=crcr, x expr=\thisrow{X}+1]{%
X	Y\\
-inf	0\\
0.5	0.512648184448801\\
1.5	0.690282833940073\\
2.5	0.797349015215159\\
3.5	0.87407822272006\\
4.5	0.924670960515262\\
5.5	0.95566134602819\\
6.5	0.974890320171754\\
7.5	0.986465042471763\\
8.5	0.99290581536451\\
9.5	0.996452907682255\\
10.5	0.997853075702418\\
11.5	0.999066554653225\\
12.5	0.99925324372258\\
13.5	0.999813310930645\\
14.5	0.999906655465323\\
15.5	0.999906655465323\\
inf	1\\
};

\addplot [color=spaceCadet, line width=0.9pt, mark=x, mark size=1.5pt]
  table[row sep=crcr, x expr=\thisrow{X}+1]{%
X	Y\\
  0 -1\\
};
\addlegendentry{\gls{wf}, $p_w = 0.1$}

\addplot [color=desireRed, line width=0.9pt, forget plot]
  table[row sep=crcr, x expr=\thisrow{X}+1]{%
X	Y\\
-inf	0\\
0	0\\
0	0.266796970437332\\
1	0.266796970437332\\
1	0.472636208160274\\
2	0.472636208160274\\
2	0.634620083068654\\
3	0.634620083068654\\
3	0.754458832152455\\
4	0.754458832152455\\
4	0.8399706816516\\
5	0.8399706816516\\
5	0.89604202296604\\
6	0.89604202296604\\
6	0.935743953090643\\
7	0.935743953090643\\
7	0.962374786220376\\
8	0.962374786220376\\
8	0.978744197410213\\
9	0.978744197410213\\
9	0.987417542145126\\
10	0.987417542145126\\
10	0.993159052040068\\
11	0.993159052040068\\
11	0.996213046665038\\
12	0.996213046665038\\
12	0.997923283655021\\
13	0.997923283655021\\
13	0.99902272172001\\
14	0.99902272172001\\
14	0.999389201075006\\
15	0.999389201075006\\
15	0.999755680430002\\
17	0.999755680430002\\
17	0.999877840215001\\
18	0.999877840215001\\
18	1\\
inf	1\\
};

\addplot [color=desireRed, line width=0.9pt, forget plot, mark=diamond, mark size=1pt, only marks]
  table[row sep=crcr, x expr=\thisrow{X}+1]{%
X	Y\\
0.5	0.266796970437332\\
1.5	0.472636208160274\\
2.5	0.634620083068654\\
3.5	0.754458832152455\\
4.5	0.8399706816516\\
5.5	0.89604202296604\\
6.5	0.935743953090643\\
7.5	0.962374786220376\\
8.5	0.978744197410213\\
9.5	0.987417542145126\\
10.5	0.993159052040068\\
11.5	0.996213046665038\\
};

\addplot [color=desireRed, line width=0.9pt, mark=diamond, mark size=1pt]
  table[row sep=crcr, x expr=\thisrow{X}+1]{%
X	Y\\
0	-1\\
};
\addlegendentry{\gls{hqf}, $p_w = 0.1$}

\addplot [color=springGreen, line width=0.9pt, forget plot]
  table[row sep=crcr, x expr=\thisrow{X}+1]{%
X	Y\\
-inf	0\\
0	0\\
0	0.291681627767804\\
1	0.291681627767804\\
1	0.56852184320766\\
2	0.56852184320766\\
2	0.766247755834829\\
3	0.766247755834829\\
3	0.886415320167564\\
4	0.886415320167564\\
4	0.950927588270497\\
5	0.950927588270497\\
5	0.982645122681029\\
6	0.982645122681029\\
6	0.994015559545183\\
7	0.994015559545183\\
7	0.997606223818073\\
8	0.997606223818073\\
8	0.999162178336326\\
9	0.999162178336326\\
9	0.999760622381807\\
10	0.999760622381807\\
10	0.999880311190904\\
11	0.999880311190904\\
11	1\\
inf	1\\
};

\addplot [color=springGreen, line width=0.9pt, forget plot, mark=o, mark size=1pt, only marks]
  table[row sep=crcr, x expr=\thisrow{X}+1]{%
X	Y\\
0.5	0.291681627767804\\
1.5	0.56852184320766\\
2.5	0.766247755834829\\
3.5	0.886415320167564\\
4.5	0.950927588270497\\
5.5	0.982645122681029\\
6.5	0.994015559545183\\
7.5	0.997606223818073\\
8.5	0.999162178336326\\
9.5	0.999760622381807\\
10.5	0.999880311190904\\
};

\addplot [color=springGreen, line width=0.9pt, mark=o, mark size=1pt]
  table[row sep=crcr, x expr=\thisrow{X}+1]{%
X	Y\\
0	0.291681627767804\\
};
\addlegendentry{\gls{pa}, $p_w = 0.1$}

\addplot [color=spaceCadet, dashdotted, line width=0.9pt]
  table[row sep=crcr, x expr=\thisrow{X}+1]{%
X	Y\\
-inf	0\\
0	0\\
0	0.732058995252964\\
1	0.732058995252964\\
1	0.884408397431016\\
2	0.884408397431016\\
2	0.949508795978981\\
3	0.949508795978981\\
3	0.978511410656716\\
4	0.978511410656716\\
4	0.991305561901861\\
5	0.991305561901861\\
5	0.996484146929658\\
6	0.996484146929658\\
6	0.998946513339933\\
7	0.998946513339933\\
7	0.999619221689133\\
8	0.999619221689133\\
8	0.999923844337827\\
9	0.999923844337827\\
9	0.999961922168913\\
10	0.999961922168913\\
10	0.999987307389638\\
11	0.999987307389638\\
11	1\\
inf	1\\
};
\addlegendentry{\gls{wf}, $p_w = 0.3$}

\addplot [color=desireRed, dashdotted, line width=0.9pt]
  table[row sep=crcr, x expr=\thisrow{X}+1]{%
X	Y\\
-inf	0\\
0	0\\
0	0.431147888818163\\
1	0.431147888818163\\
1	0.68463116203409\\
2	0.68463116203409\\
2	0.830327463045477\\
3	0.830327463045477\\
3	0.914986915623453\\
4	0.914986915623453\\
4	0.96003960676144\\
5	0.96003960676144\\
5	0.981186788316005\\
6	0.981186788316005\\
6	0.991866468632859\\
7	0.991866468632859\\
7	0.99618077657543\\
8	0.99618077657543\\
8	0.998161114647429\\
9	0.998161114647429\\
9	0.999363462762572\\
10	0.999363462762572\\
10	0.999787820920857\\
11	0.999787820920857\\
11	0.999929273640286\\
12	0.999929273640286\\
12	1\\
inf	1\\
};
\addlegendentry{\gls{hqf}, $p_w = 0.3$}

\addplot [color=springGreen, dashdotted, line width=0.9pt]
  table[row sep=crcr, x expr=\thisrow{X}+1]{%
X	Y\\
-inf	0\\
0	0\\
0	0.492599407952636\\
1	0.492599407952636\\
1	0.800144011520922\\
2	0.800144011520922\\
2	0.931594527562205\\
3	0.931594527562205\\
3	0.980958476678134\\
4	0.980958476678134\\
4	0.993999519961597\\
5	0.993999519961597\\
5	0.998879910392831\\
6	0.998879910392831\\
6	0.99959996799744\\
7	0.99959996799744\\
7	1\\
inf	1\\
};
\addlegendentry{\gls{pa}, $p_w = 0.3$}

\end{axis}

\input{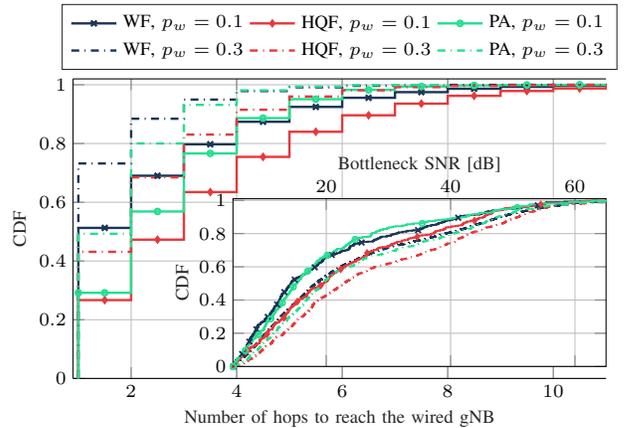}
\end{tikzpicture}%
	\caption{Comparison of \gls{wf}, \gls{hqf} and \gls{pa} policies, without \gls{wbf}, for $M=64$ antennas at the \glspl{gnb}, $\lambda_g=30$ \gls{gnb}/km$^2$ and $p_w=0.3$.}
	\label{fig:policies}
\end{figure}

\textbf{Path selection policies --}
Fig.~\ref{fig:policies} compares the performance of the different path selection algorithms presented in Sec.~\ref{sec:policies} for different values of $p_w$, without the \gls{wbf}.
In general, increasing $p_w$ makes it possible to minimize the number of hops required to forward the backhaul traffic from a wireless node to the core network and, at the same time, guarantees more efficient relaying operations. However, the trade-off oscillates between more robust  backhauling  and more expensive network deployment and  management.
Moreover, although the \gls{hqf} policy delivers the best bottleneck SNR performance, it exhibits the worst behavior in terms of number of hops,  as it greedily selects the strongest available \gls{gnb} as a relay regardless of the nature (i.e., wired or wireless) of the destination node. 
On the other hand, both \gls{wf} and \gls{pa} mechanisms have the potential to  reduce the number of hops since the  selection is biased by the availability of the wired \gls{gnb} (independent of the quality of other surrounding cells) and by context information related to the position of the wired nodes, respectively.
Conversely, both approaches degrade the quality of the bottleneck link as they may end up selecting a suboptimal  node among all the candidate relays within reach.

Interestingly, we observe that, when the number of available wired \glspl{gnb} is very low (i.e., $p_w=0.1$ and for low SNR regimes), the \gls{pa} policy performs better than the \gls{wf} in terms of both number of hops and bottleneck \gls{snr}. As can be seen in Fig.~\ref{fig:policies}, indeed, the \gls{pa} policy needs a smaller number of hops than \gls{wf} (and also \gls{hqf}) for 15\% of the paths with 4 or more hops.
In low \gls{snr} and $\lambda_{w,g}$ regimes, the \gls{wf} scheme  asymptotically operates as the \gls{hqf} and, therefore, the best choice is to select the parent which is geographically closer to a wired \gls{gnb}  with the \gls{pa} strategy.\footnote{For $p_w=0.3$ this phenomenon is obviously less pronounced but still the \gls{pa} and \gls{wf} paradigms reveal comparable performance in low SNR~regimes.}

\begin{figure*}
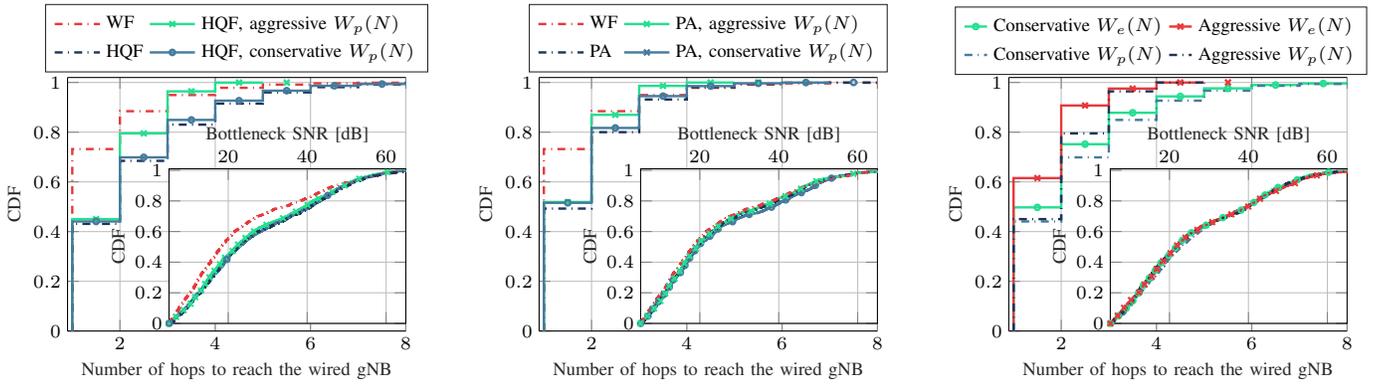

	\ifexttikz
		\tikzsetnextfilename{cdfHopsLambda30Antenna64BestWithPF}
	\fi
	\begin{subfigure}[t]{0.31\textwidth}
		\centering
		\setlength\fwidth{.8\columnwidth}
		\setlength\fheight{0.6\columnwidth}
%
%
\begin{tikzpicture}
\pgfplotsset{every tick label/.append style={font=\scriptsize}}
\begin{axis}[%
width=\fwidth,
height=\fheight,
at={(0\fwidth,0\fheight)},
scale only axis,
unbounded coords=jump,
axis y line*=left,
xmin=0.9,
xmax=8,
xlabel style={font=\scriptsize\color{white!15!black}},
xlabel={Number of hops to reach the wired gNB},
ymin=0,
ymax=1.02,
ylabel style={font=\scriptsize\color{white!15!black}},
ylabel={CDF},
axis background/.style={fill=white},
xmajorgrids,
ymajorgrids,
legend style={font=\scriptsize,legend cell align=left, align=left, draw=white!15!black,at={(0.5, 1.02)},anchor=south},
legend columns=2,
xlabel shift=-2pt,
ylabel shift=-2pt,
xticklabel shift=-1pt,
yticklabel shift=-1pt,
]

\addplot [color=desireRed, dashdotted, line width=0.9pt]
  table[row sep=crcr, x expr=\thisrow{X}+1]{%
X	Y\\
-inf	0\\
0	0\\
0	0.732058995252964\\
1	0.732058995252964\\
1	0.884408397431016\\
2	0.884408397431016\\
2	0.949508795978981\\
3	0.949508795978981\\
3	0.978511410656716\\
4	0.978511410656716\\
4	0.991305561901861\\
5	0.991305561901861\\
5	0.996484146929658\\
6	0.996484146929658\\
6	0.998946513339933\\
7	0.998946513339933\\
7	0.999619221689133\\
8	0.999619221689133\\
8	0.999923844337827\\
9	0.999923844337827\\
9	0.999961922168913\\
10	0.999961922168913\\
10	0.999987307389638\\
11	0.999987307389638\\
11	1\\
inf	1\\
};
\addlegendentry{\gls{wf}}

\addplot [color=springGreen, line width=0.9pt, forget plot]
  table[row sep=crcr, x expr=\thisrow{X}+1]{%
X	Y\\
-inf	0\\
0	0\\
0	0.450222031227618\\
1	0.450222031227618\\
1	0.795086663801748\\
2	0.795086663801748\\
2	0.964188511674545\\
3	0.964188511674545\\
3	0.999570262140095\\
4	0.999570262140095\\
4	1\\
inf	1\\
};

\addplot [color=springGreen, line width=0.9pt, forget plot, mark=x, mark size=1.5pt, only marks]
  table[row sep=crcr, x expr=\thisrow{X}+1]{%
X	Y\\
0.5	0.450222031227618\\
1.5	0.795086663801748\\
2.5	0.964188511674545\\
3.5	0.999570262140095\\
4.5	0.999570262140095\\
inf	1\\
};

\addplot [color=springGreen, line width=0.9pt, mark=x, mark size=1.5pt]
  table[row sep=crcr, x expr=\thisrow{X}+1]{%
X	Y\\
0 -1\\
};
\addlegendentry{\gls{hqf}, aggressive $W_p(N)$}

\addplot [color=spaceCadet, dashdotted, line width=0.9pt]
  table[row sep=crcr, x expr=\thisrow{X}+1]{%
X	Y\\
-inf	0\\
0	0\\
0	0.431147888818163\\
1	0.431147888818163\\
1	0.68463116203409\\
2	0.68463116203409\\
2	0.830327463045477\\
3	0.830327463045477\\
3	0.914986915623453\\
4	0.914986915623453\\
4	0.96003960676144\\
5	0.96003960676144\\
5	0.981186788316005\\
6	0.981186788316005\\
6	0.991866468632859\\
7	0.991866468632859\\
7	0.99618077657543\\
8	0.99618077657543\\
8	0.998161114647429\\
9	0.998161114647429\\
9	0.999363462762572\\
10	0.999363462762572\\
10	0.999787820920857\\
11	0.999787820920857\\
11	0.999929273640286\\
12	0.999929273640286\\
12	1\\
inf	1\\
};
\addlegendentry{\gls{hqf}}

\addplot [color=queenBlue, line width=0.9pt, forget plot]
  table[row sep=crcr, x expr=\thisrow{X}+1]{%
X	Y\\
-inf	0\\
0	0\\
0	0.440750034852921\\
1	0.440750034852921\\
1	0.698731353687439\\
2	0.698731353687439\\
2	0.84929597100237\\
3	0.84929597100237\\
3	0.92680886658302\\
4	0.92680886658302\\
4	0.966889725358985\\
5	0.966889725358985\\
5	0.986616478460895\\
6	0.986616478460895\\
6	0.993796180119894\\
7	0.993796180119894\\
7	0.997630001394117\\
8	0.997630001394117\\
8	0.999302941586505\\
9	0.999302941586505\\
9	0.999651470793252\\
10	0.999651470793252\\
10	0.999860588317301\\
11	0.999860588317301\\
11	1\\
inf	1\\
};

\addplot [color=queenBlue, line width=0.9pt, forget plot, mark=o, mark size=1pt, only marks]
  table[row sep=crcr, x expr=\thisrow{X}+1]{%
X	Y\\
0.5	0.440750034852921\\
1.5	0.698731353687439\\
2.5	0.84929597100237\\
3.5	0.92680886658302\\
4.5	0.966889725358985\\
5.5	0.986616478460895\\
6.5	0.993796180119894\\
7.5	0.997630001394117\\
8.5	0.999302941586505\\
9.5	0.999651470793252\\
10.5	0.999860588317301\\
};

\addplot [color=queenBlue, line width=0.9pt, mark=o, mark size=1pt]
  table[row sep=crcr, x expr=\thisrow{X}+1]{%
X	Y\\
  0 -1\\
};
\addlegendentry{\gls{hqf}, conservative $W_p(N)$}

\end{axis}

\input{./figures/cdfSNRLambda30Antenna64BestWithPF.tex}
\end{tikzpicture}%
		\caption{Comparison of \gls{wf}, and \gls{hqf} policies with and without \gls{wbf}.}
		\label{fig:hqf-wbf}
	\end{subfigure}\hfill
	\ifexttikz
		\tikzsetnextfilename{cdfHopsLambda30Antenna64GeoWithPF}
	\fi
	\begin{subfigure}[t]{0.31\textwidth}
		\centering
		\setlength\fwidth{.8\columnwidth}
		\setlength\fheight{0.6\columnwidth}
%
%
\begin{tikzpicture}
\pgfplotsset{every tick label/.append style={font=\scriptsize}}
\begin{axis}[%
width=\fwidth,
height=\fheight,
at={(0\fwidth,0\fheight)},
scale only axis,
unbounded coords=jump,
axis y line*=left,
xmin=0.9,
xmax=8,
xlabel style={font=\scriptsize\color{white!15!black}},
xlabel={Number of hops to reach the wired gNB},
ymin=0,
ymax=1.02,
ylabel style={font=\scriptsize\color{white!15!black}},
ylabel={CDF},
axis background/.style={fill=white},
xmajorgrids,
ymajorgrids,
legend style={font=\scriptsize,legend cell align=left, align=left, draw=white!15!black,at={(0.5, 1.02)},anchor=south},
legend columns=2,
xlabel shift=-2pt,
ylabel shift=-2pt,
xticklabel shift=-1pt,
yticklabel shift=-1pt,
]

\addplot [color=desireRed, dashdotted, line width=0.9pt]
  table[row sep=crcr, x expr=\thisrow{X}+1]{%
X	Y\\
-inf	0\\
0	0\\
0	0.732058995252964\\
1	0.732058995252964\\
1	0.884408397431016\\
2	0.884408397431016\\
2	0.949508795978981\\
3	0.949508795978981\\
3	0.978511410656716\\
4	0.978511410656716\\
4	0.991305561901861\\
5	0.991305561901861\\
5	0.996484146929658\\
6	0.996484146929658\\
6	0.998946513339933\\
7	0.998946513339933\\
7	0.999619221689133\\
8	0.999619221689133\\
8	0.999923844337827\\
9	0.999923844337827\\
9	0.999961922168913\\
10	0.999961922168913\\
10	0.999987307389638\\
11	0.999987307389638\\
11	1\\
inf	1\\
};
\addlegendentry{\gls{wf}}

\addplot [color=springGreen, forget plot, line width=0.9pt]
  table[row sep=crcr, x expr=\thisrow{X}+1]{%
X	Y\\
-inf	0\\
0	0\\
0	0.519064689587746\\
1	0.519064689587746\\
1	0.869724621150399\\
2	0.869724621150399\\
2	0.986149584487535\\
3	0.986149584487535\\
3	0.999674107870295\\
4	0.999674107870295\\
4	1\\
inf	1\\
};

\addplot [color=springGreen, forget plot, line width=0.9pt, mark=x, mark size=1.5pt, only marks]
  table[row sep=crcr, x expr=\thisrow{X}+1]{%
X	Y\\
0.5	0.519064689587746\\
1.5	0.869724621150399\\
2.5	0.986149584487535\\
3.5	0.999674107870295\\
4.5	1\\
};

\addplot [color=springGreen, line width=0.9pt, mark=x, mark size=1.5pt]
  table[row sep=crcr, x expr=\thisrow{X}+1]{%
X	Y\\
-inf	0\\
0	-1\\
};
\addlegendentry{\gls{pa}, aggressive $W_p(N)$}

\addplot [color=spaceCadet, dashdotted, line width=0.9pt]
  table[row sep=crcr, x expr=\thisrow{X}+1]{%
X	Y\\
-inf	0\\
0	0\\
0	0.492599407952636\\
1	0.492599407952636\\
1	0.800144011520922\\
2	0.800144011520922\\
2	0.931594527562205\\
3	0.931594527562205\\
3	0.980958476678134\\
4	0.980958476678134\\
4	0.993999519961597\\
5	0.993999519961597\\
5	0.998879910392831\\
6	0.998879910392831\\
6	0.99959996799744\\
7	0.99959996799744\\
7	1\\
inf	1\\
};
\addlegendentry{\gls{pa}}

\addplot [color=queenBlue, line width=0.9pt, forget plot]
  table[row sep=crcr, x expr=\thisrow{X}+1]{%
X	Y\\
-inf	0\\
0	0\\
0	0.515856069592724\\
1	0.515856069592724\\
1	0.817002767892448\\
2	0.817002767892448\\
2	0.945037564254646\\
3	0.945037564254646\\
3	0.985448793989719\\
4	0.985448793989719\\
4	0.99644128113879\\
5	0.99644128113879\\
5	0.999604586793199\\
6	0.999604586793199\\
6	1\\
inf	1\\
};

\addplot [color=queenBlue, line width=0.9pt, forget plot, only marks, mark=o, mark size=1pt]
  table[row sep=crcr, x expr=\thisrow{X}+1]{%
X	Y\\
0.5	0.515856069592724\\
1.5	0.817002767892448\\
2.5	0.945037564254646\\
3.5	0.985448793989719\\
4.5	0.99644128113879\\
5.5	0.999604586793199\\
6.5	1\\
};

\addplot [color=queenBlue, line width=0.9pt, mark=x, mark size=1.5pt]
  table[row sep=crcr, x expr=\thisrow{X}+1]{%
X	Y\\
0	-1\\
};
\addlegendentry{\gls{pa}, conservative $W_p(N)$}

\end{axis}
\input{./figures/cdfSNRLambda30Antenna64GeoWithPF.tex}
\end{tikzpicture}%
		\caption{Comparison of \gls{wf}, and \gls{pa} policies with and without \gls{wbf}.}
		\label{fig:pa-wbf}
	\end{subfigure}\hfill
	\ifexttikz
		\tikzsetnextfilename{cdfHopsLambda30Antenna64BestWithPFPolyVsExp}
	\fi
	\begin{subfigure}[t]{0.31\textwidth}
		\centering
		\setlength\fwidth{.8\columnwidth}
		\setlength\fheight{0.6\columnwidth}
%
%
\begin{tikzpicture}
\pgfplotsset{every tick label/.append style={font=\scriptsize}}
\pgfplotsset{
compat=1.11,
legend image code/.code={
\draw[mark repeat=2,mark phase=2]
plot coordinates {
(0cm,0cm)
(0.15cm,0cm)        
(0.3cm,0cm)         
};%
}
}
\begin{axis}[%
width=\fwidth,
height=\fheight,
at={(0\fwidth,0\fheight)},
scale only axis,
unbounded coords=jump,
axis y line*=left,
xmin=0.9,
xmax=8,
xlabel style={font=\scriptsize\color{white!15!black}},
xlabel={Number of hops to reach the wired gNB},
ymin=0,
ymax=1.02,
ylabel style={font=\scriptsize\color{white!15!black}},
ylabel={CDF},
axis background/.style={fill=white},
xmajorgrids,
ymajorgrids,
legend style={font=\scriptsize,legend cell align=left, align=left, draw=white!15!black,at={(0.45, 1.01)},anchor=south},
legend columns=2,
xlabel shift=-2pt,
ylabel shift=-2pt,
xticklabel shift=-1pt,
yticklabel shift=-1pt,
]
\addplot [color=springGreen, line width=0.9pt, forget plot]
  table[row sep=crcr, x expr=\thisrow{X}+1]{%
X	Y\\
-inf	0\\
0	0\\
0	0.497463668768851\\
1	0.497463668768851\\
1	0.75164518782561\\
2	0.75164518782561\\
2	0.878050452426652\\
3	0.878050452426652\\
3	0.943926514943789\\
4	0.943926514943789\\
4	0.976281875514121\\
5	0.976281875514121\\
5	0.989786125582671\\
6	0.989786125582671\\
6	0.995887030435975\\
7	0.995887030435975\\
7	0.998491911159857\\
8	0.998491911159857\\
8	0.999725802029065\\
9	0.999725802029065\\
9	0.999862901014532\\
10	0.999862901014532\\
10	1\\
inf	1\\
};

\addplot [color=springGreen, line width=0.9pt, forget plot, mark=o, mark size=1pt, only marks]
  table[row sep=crcr, x expr=\thisrow{X}+1]{%
X	Y\\
0.5	0.497463668768851\\
1.5	0.75164518782561\\
2.5	0.878050452426652\\
3.5	0.943926514943789\\
4.5	0.976281875514121\\
5.5	0.989786125582671\\
6.5	0.995887030435975\\
7.5	0.998491911159857\\
8.5	0.999725802029065\\
9.5	0.999862901014532\\
10.5	1\\
};

\addplot [color=springGreen, line width=0.9pt, mark=o, mark size=1pt]
  table[row sep=crcr, x expr=\thisrow{X}+1]{%
X	Y\\
0	-1\\
};
\addlegendentry{Conservative $W_e(N)$}

\addplot [color=desireRed, line width=0.9pt, forget plot]
  table[row sep=crcr, x expr=\thisrow{X}+1]{%
X	Y\\
-inf	0\\
0	0\\
0	0.615324803845338\\
1	0.615324803845338\\
1	0.907400862373648\\
2	0.907400862373648\\
2	0.975401145119106\\
3	0.975401145119106\\
3	0.999787940906199\\
4	0.999787940906199\\
4	1\\
inf	1\\
};

\addplot [color=desireRed, line width=0.9pt, forget plot, mark=x, mark size=1.5pt, only marks]
  table[row sep=crcr, x expr=\thisrow{X}+1]{%
X	Y\\
0.5	0.615324803845338\\
1.5	0.907400862373648\\
2.5	0.975401145119106\\
3.5	0.999787940906199\\
4.5	1\\
};

\addplot [color=desireRed, line width=0.9pt, mark=x, mark size=1.5pt]
  table[row sep=crcr, x expr=\thisrow{X}+1]{%
X	Y\\
0	-1\\
};
\addlegendentry{Aggressive $W_e(N)$}

\addplot [color=queenBlue, dashdotted, line width=0.9pt]
  table[row sep=crcr, x expr=\thisrow{X}+1]{%
X	Y\\
-inf	0\\
0	0\\
0	0.440750034852921\\
1	0.440750034852921\\
1	0.698731353687439\\
2	0.698731353687439\\
2	0.84929597100237\\
3	0.84929597100237\\
3	0.92680886658302\\
4	0.92680886658302\\
4	0.966889725358985\\
5	0.966889725358985\\
5	0.986616478460895\\
6	0.986616478460895\\
6	0.993796180119894\\
7	0.993796180119894\\
7	0.997630001394117\\
8	0.997630001394117\\
8	0.999302941586505\\
9	0.999302941586505\\
9	0.999651470793252\\
10	0.999651470793252\\
10	0.999860588317301\\
11	0.999860588317301\\
11	1\\
inf	1\\
};
\addlegendentry{Conservative $W_p(N)$}

\addplot [color=spaceCadet, dashdotted, line width=0.9pt]
  table[row sep=crcr, x expr=\thisrow{X}+1]{%
X	Y\\
-inf	0\\
0	0\\
0	0.450222031227618\\
1	0.450222031227618\\
1	0.795086663801748\\
2	0.795086663801748\\
2	0.964188511674545\\
3	0.964188511674545\\
3	0.999570262140095\\
4	0.999570262140095\\
4	1\\
inf	1\\
};
\addlegendentry{Aggressive $W_p(N)$}

\end{axis}

\input{./figures/cdfSNRLambda30Antenna64BestWithPFPolyVsExp.tex}
\end{tikzpicture}%
		\caption{Comparison \gls{hqf} policies with polynomial and exponential \gls{wbf}.}
		\label{fig:polyexp}
	\end{subfigure}
\caption{Impact of the \gls{wbf} (aggressive or conservative, polynomial or exponential) on the performance.}
\label{fig:wbf}
\end{figure*}

\textbf{\gls{wbf}  configurations --}
In Fig.~\ref{fig:hqf-wbf} we compare the behavior of the \gls{hqf} and the \gls{wf} policies when  considering different \gls{wbf} configurations to bias the path selection results.
First, we see that, since the  \gls{wf} approach is designed to minimize the number of hops to reach a wired \gls{gnb}, it generally outperforms any other architecture for the hop-count metric. 
However,  the quality of the bottleneck link  inevitably decreases (on average by more than 4 dB compared to its \gls{hqf} counterpart), thereby increasing the risk of communication outage between the endpoints. 
Moreover,  for bad SNR regimes (i.e., as the probability of detecting valid wired nodes reduces) the   \gls{hqf} scheme implementing aggressive \gls{wbf} achieves the best performance in terms of both number of hops and bottleneck SNR.

Second, we observe that, albeit a conservative \gls{wbf} applied to an \gls{hqf} scheme does not provide any significant  performance improvements with respect to a pure \gls{hqf} approach, a more aggressive design of the bias function has the ability to remarkably reduce the number of hops required to forward the backhaul traffic to a wired \gls{gnb}, without any visible degradation in terms of \gls{snr}.
We deduce that it is highly convenient to configure very aggressive\footnote{Of course, if the \gls{wbf} parameters are too aggressively configured, the \gls{hqf} approach will more likely operate as a \gls{wf} policy, with all that this implies (including, but not limited to, a detrimental degradation of the bottleneck~SNR).} \gls{wbf} functions since, for a multi-hop scenario, they deliver more efficient relaying operations without affecting the quality of the communication. 

The same conclusions can be drawn by comparing the performance of the \gls{pa} and the \gls{wf} policies as a function of the  different \gls{wbf} configurations.
In this regard, Fig.~\ref{fig:pa-wbf} illustrates how the biased \gls{pa} approach guarantees very fast and high-quality backhauling thanks to the  low number of hops that needs to be made before successfully forwarding the traffic to the core network and the relatively large bottleneck SNR that is experienced. In particular, the reduction in the number of hops is even beyond the capabilities of the biased \gls{hqf} counterpart, at the cost of a slight reduction in the bottleneck \gls{snr} (in the order of 2 dB on the 50\% percentile). 
Moreover, as already mentioned before, both biased and unbiased \gls{pa} architectures outperform the \gls{wf} scheme in the case of low \gls{snr} regimes.

Finally, in Fig.~\ref{fig:polyexp} we compare the behavior of the \gls{hqf} policy with polynomial and exponential \glspl{wbf}.
Based on the design choices presented in Tab.~\ref{table:wbf} and according to Eqs.~\eqref{eq:Wp} and \eqref{eq:We}, the exponential bias function is more aggressive than the polynomial one for all values of $N$, i.e., the current number of hops.
However, the exponentially-biased  \gls{hqf} approach, because of its inherently aggressive nature, is affected by SNR  deterioration, though moderate (i.e., smaller than 1 dB on average), with respect to its polynomially-biased counterpart.

\begin{figure}[t]
	\centering
	\setlength\fwidth{0.8\columnwidth}
	\setlength\fheight{0.45\columnwidth}
%
%
\begin{tikzpicture}
\pgfplotsset{every tick label/.append style={font=\scriptsize}}
\begin{axis}[%
width=\fwidth,
height=\fheight,
at={(0\fwidth,0\fheight)},
scale only axis,
unbounded coords=jump,
axis y line*=left,
xmin=0.9,
xmax=8,
xlabel style={font=\scriptsize\color{white!15!black}},
xlabel={Number of hops to reach the wired gNB},
ymin=0,
ymax=1.02,
ylabel style={font=\scriptsize\color{white!15!black}},
ylabel={CDF},
axis background/.style={fill=white},
xmajorgrids,
ymajorgrids,
legend style={font=\scriptsize,legend cell align=left, align=left, draw=white!15!black,at={(0.5, 1.01)},anchor=south},
xlabel shift=-2pt,
ylabel shift=-2pt,
xticklabel shift=-1pt,
yticklabel shift=-1pt,
]
\addplot [color=queenBlue, dashdotted, line width=0.9pt]
  table[row sep=crcr, x expr=\thisrow{X}+1]{%
X	Y\\
-inf	0\\
0	0\\
0	0.431903897192345\\
1	0.431903897192345\\
1	0.697024724123481\\
2	0.697024724123481\\
2	0.84718536108395\\
3	0.84718536108395\\
3	0.925827629557201\\
4	0.925827629557201\\
4	0.968850398100293\\
5	0.968850398100293\\
5	0.985752200027937\\
6	0.985752200027937\\
6	0.994552311775388\\
7	0.994552311775388\\
7	0.998044419611678\\
8	0.998044419611678\\
8	0.999441262746194\\
9	0.999441262746194\\
9	1\\
inf	1\\
};
\addlegendentry{\gls{mlr} policy, conservative $W_p(N)$}

\addplot [color=springGreen, line width=0.9pt]
  table[row sep=crcr, x expr=\thisrow{X}+1]{%
X	Y\\
-inf	0\\
0	0\\
0	0.453554295682512\\
1	0.453554295682512\\
1	0.782671899985463\\
2	0.782671899985463\\
2	0.963221398459078\\
3	0.963221398459078\\
3	0.99941852013374\\
4	0.99941852013374\\
4	1\\
inf	1\\
};
\addlegendentry{\gls{mlr} policy, aggressive $W_p(N)$}

\addplot [color=desireRed, line width=0.9pt, dashed]
 table[row sep=crcr, x expr=\thisrow{X}+1]{%
X	Y\\
-inf	0\\
0	0\\
0	0.331331191507194\\
1	0.331331191507194\\
1	0.551753038133818\\
2	0.551753038133818\\
2	0.705266098617125\\
3	0.705266098617125\\
3	0.814778600363179\\
4	0.814778600363179\\
4	0.88559854728314\\
5	0.88559854728314\\
5	0.928341947199329\\
6	0.928341947199329\\
6	0.959491549099036\\
7	0.959491549099036\\
7	0.977231456907389\\
8	0.977231456907389\\
8	0.987009358849001\\
9	0.987009358849001\\
9	0.993574521581226\\
10	0.993574521581226\\
10	0.996647576477162\\
11	0.996647576477162\\
11	0.998184103925129\\
12	0.998184103925129\\
12	0.99916189411929\\
13	0.99916189411929\\
13	0.999580947059645\\
14	0.999580947059645\\
14	0.999720631373097\\
15	0.999720631373097\\
15	1\\
inf	1\\
};
\addlegendentry{\gls{mlr} policy}

\end{axis}

\input{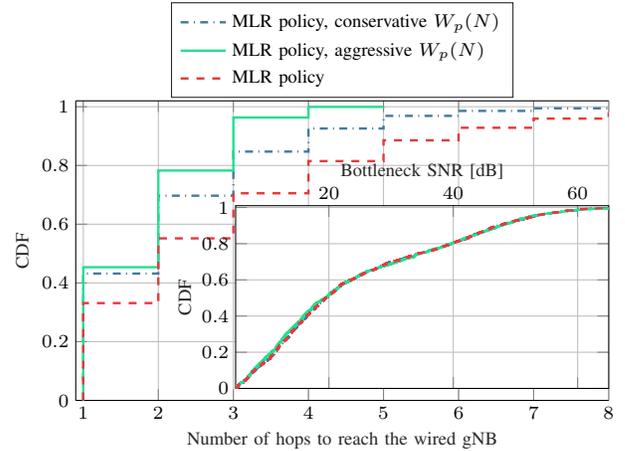}
\end{tikzpicture}%
	\caption{Comparison of \gls{mlr} policy with and without \gls{wbf}.}
	\label{fig:rate}
\end{figure}

\textbf{\gls{mlr} performance --}
While the \gls{iab} results presented in the previous paragraphs were based on \gls{snr} considerations, i.e., the candidate parent is chosen according to the instantaneous quality of the received signal, the \gls{cdf} curves displayed in Fig.~\ref{fig:rate} analyze the performance of the \gls{mlr} backhauling approach which relies on the instantaneous cell load and the Shannon rate as a metric for the path selection operations.
We observe that  Fig.~\ref{fig:rate}  leads to the same conclusions previously set out, i.e., the design of aggressive polynomial bias functions has the potential to significantly reduce the number of hops without affecting the quality of the communication (in terms of bottleneck SNR).
Aggressive exponential \glspl{wbf} are able to further reduce the number of relaying events, though this may slightly undermine the quality of the weakest link.

\ifexttikz
	\tikzsetnextfilename{cdfHopsLambda30Antenna64RateWithPF}
\fi

\subsection{Final Considerations} 
\label{sub:final_considerations}
Based on the above discussion, in the following we provide guidelines on how to optimally configure the path selection policies presented in the previous sections to maximize the performance of the \gls{iab} traffic relaying operations.

We state that a \gls{wf} approach, although minimizing the number of hops required to connect to a wired \gls{gnb}, is affected by performance degradation in terms of bottleneck SNR.
Moreover, this scheme has been proven  particularly inefficient when reducing the number of wired nodes (i.e., for low values of $p_w$) and for low SNR regimes (i.e., when configuring very wide beams and considering sparsely deployed networks).

In this context, a \gls{pa} strategy  may deliver improved performance leveraging on context information (e.g., the position of the surrounding wired \glspl{gnb}) that is periodically distributed throughout the network.
Furthermore, it is possible to design  aggressive polynomial and  exponential  \glspl{wbf} to bias the relay selection procedures and further reduce the overall number of hops without significant performance degradation in the quality of the weakest link.

\section{Conclusions and Future Work}
\label{sec:concl}
Motivated by the fact that the integration between mmWave access and wireless backhaul is becoming a reality, 
in this paper we compared the performance of different distributed path selection strategies to efficiently forward the backhaul traffic (possibly through multiple hops) from a wireless gNB to a wired gNB connected to the core network.
The investigated policies may or may not leverage on a function that biases the  link selection towards base stations with wired backhaul capabilities, to  minimize the latency of the relaying operations.
We showed through simulations that it is always possible to decrease the number of hops required to connect to a wired gNB by designing aggressive bias functions without affecting the average bottleneck SNR (i.e., the quality of the weakest link).
Moreover, we demonstrated that a WF strategy  that always selects a wired gNB as the next hop, if available, is ineffective in case of sparsely deployed networks and for low SNR regimes.
Conversely, a PA scheme which uses context information to perform the path selection has the potential to significantly improve the  overall performance  in terms of both number of hops and communication quality.

As part of our future work, we will set up ns--3 based  simulations to evaluate the end-to-end performance of the presented  \gls{iab} policies in terms of experienced throughput, latency, and packet loss ratio, and considering realistic traffic~models.

\bibliographystyle{IEEEtran}
\bibliography{bibl.bib}

\end{document}